\documentclass[11pt]{article}
\usepackage[latin1]{inputenc}
\usepackage[english]{babel}
\usepackage{graphicx}
\usepackage[namelimits]{amsmath}
\usepackage{amssymb}
\usepackage{amsmath}
\usepackage{amsthm}

\begin{document}
\title{A novel derivation for Kerr  metric in Papapetrou  gauge}
\author{Roberto Bergamini\\
Istituto di radioastronomia I.R.A. , C.N.R, Bologna-Italy,\\
Stefano Viaggiu
\\
Dipartimento di Matematica, Universit\'a di Roma ``Tor Vergata'',\\
Via della Ricerca Scientifica, 1, I-00133 Roma, Italy\\
E-mail: viaggiu@mat.uniroma2.it\\
(or: stefano.viaggiu@ax0rm1.roma1.infn.it)}
\date{\today}\maketitle
\begin{abstract}
We present a simple novel derivation, ab initio, of the equations
appropriate for  stationary axisymmetric spacetimes using
the Papapetrou form of the metric (Papapetrou  gauge).
It is shown that using coordinates 
which preserve the Papapetrou gauge 
three separated solutions of the
Ernst equations appear in the case of Kerr metric.
In this context a parameter arises which represents topological
defects induced by an infinite static string along the z axis.
Finally, we discuss a simple solution that may be 
derived from the Kerr ansatz.
\end{abstract}
PACS numbers: 04.20.-q, 04.20.Jb\\
Journal: {\it Class. Quantum Grav.} {\bf 21} (2004) 4567-4573.
\section{Introduction}
Contrary to the claim of Landau and Lifshitz 
that ``there is no constructive analytic derivation of the Kerr metric
that is adequate in its physical ideas and even a check of this
solution of Einstein's equations involves 
cumbersome calcualtions'' \cite{L1},
Chandrasekhar \cite{chandra} first derived and properly reduced the
equations leading to the Kerr metric. He used the spheroidal coordinate
$\eta$ by setting $r=m+\sqrt{m^2-a^2}\;\eta$ (here $m$ is the mass and
$ma$ is the angular momentum), instead of the much more commonly used
Boyer-Lindquist (BL)
coordinate $r$ \cite{BL1}.
Although allowing a presentation of the
metric in a simple (polynomial) form, this coordinate system
leads to a confusing
interpretation of the radial variable $r$ (see \cite{chandra}, pg. 341,357).
The Kerr metric is also commonly derived
by using a method introduced by Ernst \cite{ernst}.
Namely, one shifts to prolate coordinates and again obtains a polynomial form
characterized by the relation $m^2-a^2=1$ 
(notice this implies $a^2<m^2$) and recovers the standard expression
performing a transformation to the BL coordinates.
It is clear that this simple procedure does not 
allow the treatment of the case $a^2>m^2$ but, 
perhaps less obviously, even the limit $q\rightarrow 1$ (here $q=a/m$),
corresponding to the so called extreme ($a^2=m^2$) 
Kerr solution, cannot be described in a physically satisfactory way. 
As a matter of fact in the literature 
(consult for example \cite{TS} and references therein) 
the following relation between cylindrical and
spheroidal prolate coordinates is exploited:
$\rho=m\sqrt{1-q^2}\sqrt{x^2-1}\sqrt{1-y^2},\;z=m\sqrt{1-q^2}\; xy$. 
This amounts to considering the unit lenght as
depending on $a,m$. As a consequence, for the 
coordinate transformation quoted above to make 
sense when $q\rightarrow 1$, we must send $x\rightarrow \infty$.
To put it in a somewhat fancy manner: in order 
to ``see the Kerr metric'' the observer must sit at the point at infinity.
Moreover after this procedure we are left with no 
information about the requirements needed to obtain 
the extreme solution in the same spirit as the static one, namely by
continuously modifying the parameters  without
changing the coordinates.
This paper tries to overcome
these problems.\\ 
The Papapetrou gauge \cite{Pap} 
permits a description of the Kerr line element 
avoiding these shortcomings and we will 
show that in this case three distinct solutions 
appear that cannot be transfomed into each other 
by continuously varying the two parameters $a$ and $m$.
In section 2 we perform a novel derivation of the 
field equations while in section 3 we study the 
covariance of the Ernst equations
for general transformation of coordinates. 
In particular we observe that in the Papapetrou gauge 
they are invariant in form under analytic coordinate 
transformations thereby proving that these latter (and only these) 
leave the line element invariant in form. In section 4 we write down the
three different Kerr solutions and use this property to prove 
that no continuous variation of the parameters 
connect them (without changing the gauge). 
In Section 5 we study the Kerr solution in presence of a cosmic string.
Section 6 collects some conclusions.
\section{Derivation of the field equations for the Kerr metric}
Our starting point is the 
axially symmetric line element in the form
\begin{equation}
ds^2=e^{\nu}\left[{(dx^1)}^2+{(dx^2)}^2\right]+l{d\varphi}^2+
 2m d\varphi dt-f{dt}^2,
\label{b1}
\end{equation}
where
$\nu=\nu(x^1 , x^2)\;,\; l=l(x^1 , x^2)\;,\;m=m(x^1 , x^2)\;,\;
f=f(x^1 , x^2)$, 
$\varphi$ is an angular coordinate , $t$ the  time coordinate,
$x^1 , x^2$ spatial coordinates and
\begin{equation}
fl + m^2 = {\rho}^2,
\label{b2}
\end{equation}
where $\rho$ is the radius in a cylindrical coordinate
system. The Ricci tensor $R_{\mu\nu}$ is defined in terms of 
Christoffel symbols by
$R_{\mu\beta}=-{\Gamma}_{\beta\mu,\alpha}^{\alpha}+
{\Gamma}_{\beta\alpha,\mu}^{\alpha}+
{\Gamma}_{\beta\alpha}^{\sigma}{\Gamma}_{\sigma\mu}^{\alpha}-
{\Gamma}_{\beta\mu}^{\sigma}{\Gamma}_{\sigma\alpha}^{\alpha}$.
For the metric (\ref{b1}) the non
identically vanishing Einstein's equations are: 
$R_{11}=R_{22}=R_{12}=R_{33}=R_{34}=R_{44}=0$ 
(with coordinates $x^1,x^2,x^3=\varphi,x^4=t$).
Thanks to (\ref{b2}) we can eliminate $l$ and choose a
coordinate system such that 
${\rho}_{11}+{\rho}_{22}=\Delta\rho =0$.
We introduce
a new function $\gamma$ by $e^{2\gamma}=fe^{\nu}$ and the function
$\omega$ by $\omega=\frac{m}{f}$. The Einstein's equations thus obtained are
\begin{eqnarray}
& &{\gamma}_1=-\frac{\Sigma {\rho}_1+\Pi{\rho}_2}
{4\rho({{\rho}_1}^2+{{\rho}_2}^2)}+\frac{c}{2}\;,\;
{\gamma}_2=\frac{\Sigma {\rho}_2-\Pi {\rho}_1}
{4\rho({{\rho}_1}^2+{{\rho}_2}^2)}+\frac{d}{2},\label{bin}\\ 
& &{\nabla}^2 f+\frac{f}{{\rho}^2}\left({{\omega}_{\alpha}}^2 f^2-
\frac{{\rho}^2}{f^2}{f_{\alpha}}^2\right)=0\;,\; 
{\tilde{\nabla}}^2 \omega+2{\omega}_{\alpha}\frac{f_{\alpha}}{f}=0. 
\label{bam}
\end{eqnarray}
A summation over $\alpha$ is implicit in equations (\ref{bam})
with $\alpha=1,2$, i.e. $x^1,x^2$, low indices denote partial derivatives
and ${\nabla}^2={\partial}^{2}_{\alpha\alpha}+\frac{{\rho}_{\alpha}}{\rho}
{\partial}_{\alpha}$, ${\tilde{\nabla}}^2={\partial}^{2}_{\alpha\alpha}-
\frac{{\rho}_{\alpha}}{\rho}{\partial}_{\alpha}$,
with
\begin{eqnarray}
& &c=\frac{[2{\rho}_{12}{\rho}_2+({\rho}_{11}-{\rho}_{22}){\rho}_1]}
{({\rho}_{1}^2+{\rho}_{2}^2)}\;,\,
d=\frac{[2{\rho}_{12}{\rho}_1-({\rho}_{11}-{\rho}_{22}){\rho}_2]}
{({\rho}_{1}^2+{\rho}_{2}^2)}, \label{And} \\
& &\Sigma=-\frac{{\rho}^2}{f^2}(f_{1}^2-f_{2}^2)+f^2
\left[\left\{\left(\frac{m}{f}\right)_1\right\}^2-
\left\{\left(\frac{m}{f}\right)_2\right\}^2\right],\nonumber\\
& &\Pi=-2{\rho}^2\frac{f_1 f_2}{f^2}+
2f^2{\left(\frac{m}{f}\right)}_1{\left(\frac{m}{f}\right)}_2.\nonumber
\end{eqnarray}
In order to write the field equations 
in a complex form, some care must be taken. 
In fact, if we define \cite{ernst} a potential ${\tilde{\phi}}$ such that
\begin{equation}
{\omega}_1 =-\frac{\rho}{f^2}{\tilde{\phi}}_2\;,\;
{\omega}_2 = \frac{\rho}{f^2}{\tilde{\phi}}_1,
\label{b77}
\end{equation}
the equation (\ref{bam}) involving 
${\tilde{\nabla}}^2\omega$ becomes an identity
and therefore cannot be taken as a field equation.\\
In any case, another equation can be drawn from the condition
${\phi}_{12} = {\phi}_{21}$ or ${\omega}_{12} = {\omega}_{21}$.
The new equations for $f$ and $\tilde{\phi}$ are 
\begin{equation} 
{\nabla}^2 f-\frac{1}{f}({f_{\alpha}}^2-
{\tilde{\phi}}_{\alpha}^2)=0\;,\;
{\nabla}^2 {\tilde{\phi}}-
\frac{2}{f}f_{\alpha}{\tilde{\phi}}_{\alpha}=0. 
\label{b89}
\end{equation}
Once equations (\ref{b89}) are solved, the other functions
$\gamma$ and $\omega$ can be obtained by a simple integration.\\
Finally, we define a complex function $\xi$, with $f+\imath{\tilde{\phi}}=
\frac{\xi-1}{\xi+1}$. It follows that
equations (\ref{b89}) can be reduced to a single complex
equation given by
\begin{equation}
\left(\xi{\xi}^{*}-1\right)\left({\xi}_{\alpha\alpha}+
\frac{{\rho}_{\alpha}}{\rho}{\xi}_{\alpha}\right)=
\left(\xi{\xi}^{*}-1\right){\nabla}^2{\xi}=
2{\xi}^{*}{{\xi}_{\alpha}}^2,
\label{b112}
\end{equation}
where ``$*$''  denotes
complex conjugation.
Summarizing, we have expressions (\ref{And}) and 
equations (\ref{b77}), (\ref{b112}) and (\ref{bin}), with 
\begin{eqnarray}
& &\Sigma =-4{\rho}^2\frac{({\xi}_1{\xi}_{1}^{*}-
{\xi}_{2}{\xi}_{2}^{*})}
{{({\xi}{\xi}^{*}-1)}^2}\;,\;
\Pi =-4{\rho}^2\frac{({\xi}_{1}^{*}{\xi}_2+{\xi}_1{\xi}_{2}^{*})}
{{(\xi{\xi}^{*}-1)}^2}, \label{b132} \\
& &f =\frac{\xi{\xi}^{*}-1}
{(\xi+1)({\xi}^{*}+1)}\;,\;
\tilde{\phi} =-\imath\frac{(\xi-{\xi}^{*})}
{(\xi+1)({\xi}^{*}+1)}. \label{b139}
\end{eqnarray}
\section{Covariance of Ernst equations for a general coordinate transformation}
In what follows we consider the behaviour of the Ernst equations
(\ref{b89}) under a general coordinate transformation.
Starting from the functions
$f=f(x,y)$ and ${\tilde{\phi}}={\tilde{\phi}}(x,y)$ we will choose
a generic coordinate transformation $x=x(x^{\prime},y^{\prime})$
and $y=y(x^{\prime},y^{\prime})$. 
After simple but very tedious calculations, we find that in the new
coordinates $x^{\prime}$, $y^{\prime}$ the equations (\ref{b89})
become
\begin{equation}
f{\nabla}^2 f(x^{\prime}, y^{\prime})-f_{A}^2+
{\tilde{\phi}}_{A}^{2}=0,\;,\;
f{\nabla}^2\tilde{\phi}(x^{\prime}, y^{\prime})-2
{\tilde{\phi}}_{A}f_{A}=0\;\;,\;\;\;A=x^{\prime}, y^{\prime}
\label{ceres}
\end{equation}
if and only if the following conditions are imposed
\begin{equation}
A=B\;,\;C=0\;,\;\Delta x^{\prime} = \Delta y^{\prime} = 0,
\label{b205}
\end{equation}
$A={\left(\frac{\partial x^{\prime}}{\partial x}\right)}^2+
     {\left(\frac{\partial x^{\prime}}{\partial y}\right)}^2\;,\;
B={\left(\frac{\partial y^{\prime}}{\partial x}\right)}^2+
     {\left(\frac{\partial y^{\prime}}{\partial y}\right)}^2\;,\;
C=\frac{\partial x^{\prime}}{\partial x}
\frac{\partial y^{\prime}}{\partial x}+
\frac{\partial x^{\prime}}{\partial y}
\frac{\partial y^{\prime}}{\partial y}$.\\
It is easy to see that the solution of the system (\ref{b205}),
thanks to 
$\frac{{\partial}^2 x^{\prime}}{\partial x\partial y}=
\frac{{\partial}^2 x^{\prime}}{\partial y\partial x}\;,\;
\frac{{\partial}^2 y^{\prime}}{\partial x\partial y}=
\frac{{\partial}^2 y^{\prime}}{\partial y\partial x}$,
is given by the harmonic ansatz
\begin{equation}
\frac{\partial x^{\prime}}{\partial x}=
\frac{\partial y^{\prime}}{\partial y}\;,\;
\frac{\partial x^{\prime}}{\partial y}=-
\frac{\partial y^{\prime}}{\partial x}. 
\label{b208}
\end{equation}
The condition (\ref{b208}) can be
simply integrated provided that
$x+\imath y = F(x^{\prime}+\imath y^{\prime})$,
where  $F$ is  analytic. Therefore, to preserve the Papapetrou gauge,
we must use analytic coordinate transformations. In the following section we
will derive the Kerr solutions using the Papapetrou gauge.
\section{Kerr solutions}
\subsection{Kerr solution with $a^2<m^2$}
Let us take spheroidal prolate coordinates defined in 
terms of the cylindrical ones by the analytic transformation
\begin{equation}
\rho = \sinh\mu\sin\theta\;,\;z=\cosh\mu\cos\theta.
\label{b140b}
\end{equation}
In this adapted coordinate system we have $ x^1=\mu\;,\;x^2=\theta\;,\;
x^3=\varphi\;,\;x^4=t$ with the line element
$ds^2=f^{-1}\left[e^{2\gamma}\left(d{\mu}^2+d{\theta}^2\right)+
{\rho}^2 d{\varphi}^2\right]-
f{\left(dt-\omega d\varphi\right)}^2$.
It is easy to see that the complex function
$\xi = p\cosh\mu+\imath q\cos\theta$
is a solution of (\ref{b112}) with $p$ and $q$ real constants satisfying
$p^2+q^2=1$.
The metric functions $f$, $\gamma$ and $\omega$ are given by  
\begin{eqnarray}
& &f=\frac{p^2{\cosh}^2\mu+q^2{\cos}^2\theta-1}
{{\left(p\cosh\mu+1\right)}^2+q^2{\cos}^2\theta}\;,\; 
\omega=2\frac{q}{p}\frac{(p\cosh\mu+1){\sin}^2\theta}
{\left[p^2{\cosh}^2\mu-1+q^2{\cos}^2\theta\right]}, \nonumber \\
& &e^{2\gamma} =\frac{\left(p^2{\cosh}^2\mu-1+q^2{\cos}^2\theta\right)}{k^2}.
\label{b188bb}
\end{eqnarray}
The integration constant $k$ is a new parameter that will be discussed later.
At this stage we set $k^2=p^2$ in equation (\ref{b188bb}) and
define $p$ and $q$ as $p=1/m$ and $q=a/m$, with
$m^2-a^2=1$.
Performing an expansion of the metric at large distances 
one sees clearly that $m$ can be identified with the mass 
and $ma$ with the angular momentum
of the source.  
Introducing the BL coordinates, defined in terms of the cylindrical ones by
\begin{equation}  
\rho =\sqrt{r^2+a^2-2mr}\;\sin\theta\;,\;z = (r-m)\cos\theta, 
\label{b190bb}
\end{equation}
(the relation between $r$ and  $\mu$ is given by
$r=\sqrt{m^2-a^2}\;\cosh\mu+m$),
the line element becomes
\begin{eqnarray}
ds^2 &=&(r^2+a^2{\cos}^2\theta)\left(d{\theta}^2+
\frac{dr^2}{r^2+a^2-2mr}\right)-dt^{2}+\nonumber \\
&+&(r^2+a^2){\sin}^2\theta d{\varphi}^{2}+
\frac{2mr}{r^2+a^2{\cos}^2\theta}
{\left(dt+a{\sin}^2\theta d\varphi\right)}^2. \label{b192bb}
\end{eqnarray}
The static limit is obtained from solution (\ref{b188bb}) by setting
$a=0$ and then $q=0$ and $p=1$ ($m=1$). This is 
consistent with physical requirements
because the parameters of the source are changed continuously.
To obtain the extreme Kerr solution
we must instead take the limit $a^2=m^2$ and 
then $q=1$, $p=0$. But then, since
$p=\frac{1}{m}$, $m^2-a^2=1$ and therefore $m^2\neq a^2$ provided that
$m,a\neq\infty$. Besides, expression (\ref{b188bb}) for $\gamma$ (with
$k^2=p^2$) diverges as $p\rightarrow 0$ and so does $\omega$ .
We can also cast $p$ and $q$ in the form
$p=\frac{\sqrt{m^2-a^2}}{m}\;,\;q=\frac{a}{m}$.
In this way, only a change in unit lenght occurs: distances
are no longer measured in units $\sqrt{m^2-a^2}$.
We have again the relation
(\ref{b190bb}) but
in expression (\ref{b140b}) $\rho$ is 
substituted by ${\rho}/\sqrt{(m^2-a^2)}$, 
$z$ by $z/\sqrt{(m^2-a^2)}$ and $k^2=\frac{1}{m^2}$.
Also, $\omega\rightarrow\sqrt{m^2-a^2}\omega$.
Owing to this fact, expression (\ref{b140b}) is not defined as
$a^2\rightarrow m^2$ and in this limit the extreme Kerr solution
does not emerge. 
On the contrary, thanks to (\ref{b192bb}), we can provide a unified description
of the Kerr solutions.
The coordinate transformation (\ref{b190bb}) is not 
analytic and the Papapetrou
gauge breaks down.
The condition $p^2+q^2=1$ fixes
the solution to be confined in the range $a^2<m^2$.
This fact is coordinate-independent, provided that the
Papapetrou  gauge is used. To show this,
we use the relation 
(\ref{b140b}) with the inverse given by
$\cosh\mu =\frac{1}{2}\left[A+B\right]\;,\;\cos\theta 
=\frac{1}{2}\left[A-B\right]$ with
$A=\sqrt{{\left(z+1\right)}^2+{\rho}^2}\;,\;
B=\sqrt{{\left(z-1\right)}^2+{\rho}^2}$.
The line element in cylindrical coordinates becomes
\begin{equation}
ds^2=f^{-1}\left[\frac{e^{2\gamma}}{AB}\left(d{\rho}^2+dz^2\right)+
{\rho}^2 d{\varphi}^2\right]-f{\left(dt-\omega d\varphi\right)}^2,
\label{b196bb}
\end{equation}
where $f$, $\omega$ and $\gamma$ are given by equations
(\ref{b188bb}) with the functions $\cosh\mu$ and $\cos\theta$ written
in terms of $\rho$ and $z$.\\
It follows that asking for the conservation of the Papapetrou gauge makes 
a unified description of the Kerr solution not achievable: 
three different ans$\ddot{\mbox{a}}$tze are 
needed and it is impossible to join one
to the other by a continuous variation of
the parameters caracterizing the source.
No such phenomenon occurs in the usual approach to the 
problem (see for example \cite{TS}) which relies on a 
relation between spheroidal prolate coordinates and 
cylindrical ones. More explicitly, first one sets 
$\rho=mp \sinh\mu\sin\theta, z=mp \cosh\mu\cos\theta$ 
($p=\frac{\sqrt{m^2-a^2}}{m}$) and then sends $p$ to 
zero and $\mu$ to infinity. It is in this way that 
the extreme Kerr solution emerges. We now 
proceed to show how this result can be obtained in our chosen gauge.
\subsection{Kerr solution with $a^2=m^2$}
To use the Papapetrou form of the metric we have 
to modify the usual spherical coordinates 
$x^1=r$, $x^2=\vartheta$, $x^3=\varphi , 
x^4=t$ with $\rho=r\sin\vartheta\;,\;z=r\cos\vartheta\;$.
In fact, we will take 
$r=e^{\nu}$ with $-\infty<\nu<\infty$ so that
$\rho = e^{\nu}\sin\vartheta$ and $z = e^{\nu}\cos\vartheta$. 
It is a simple matter to verify that the complex function
$\xi =pe^{\nu}+\imath q\cos\vartheta$ satisfies (\ref{b112}) 
with the condition
$q^2=1$.
In the same way the complex function $\xi=p\cos\vartheta+\imath qe^{\nu}$ 
is a solution
with $p^2=1$: it will be discussed in the Appendix.
We emphasize that the condition $q^2=1$ is different 
from the appropriate one for the solution with $a^2<m^2$ ( $p^2+q^2=1$ ).
Once the field equations are solved, the metric can again 
be expressed in spherical coordinates. The result is
\begin{eqnarray}
& &ds^2=\frac{1}{k^2}\left[p^2+\frac{1}{r^2}+\frac{2p}{r}+
\frac{{\cos}^2\vartheta}{r^2}\right]dr^2+
\frac{1}{k^2}\left[p^2 r^2+1+2pr+{\cos}^2\vartheta\right]
d{\vartheta}^2+ \nonumber \\
& &+\frac{\left[p^2 r^2+2pr+1+{\cos}^2\vartheta\right]
r^2{\sin}^2\vartheta}{\left(p^2 r^2-{\sin}^2\vartheta\right)}d{\varphi}^2
-\nonumber \\
& &-\frac{\left(p^2 r^2-{\sin}^2\vartheta\right)}
{\left(p^2 r^2+1+2pr+{\cos}^2\vartheta\right)}
{\left[dt-\frac{2q}{p}\frac{(pr+1){\sin}^2\vartheta}
{\left(p^2 r^2-{\sin}^2\vartheta\right)}d\varphi\right]}^2.
\label{b516}
\end{eqnarray}
Note that in these coordinates the point $r=0$ represents
the location of two coinciding horizons, i.e. the
origin of the spatial coordinates.
\subsection{Kerr solution with $a^2>m^2$}
We consider spheroidal oblate coordinates 
$\rho = \cosh\mu\cos\theta\;,\;z=\sinh\mu\sin\theta$,
with inverse given by
$\cosh\mu = \frac{1}{2}[A+B]\;,\;\cos\theta =\frac{1}{2}[A-B]$ where
$A=\sqrt{z^2+{\left(\rho +1\right)}^2}\;,\;
   B=\sqrt{z^2+{\left(\rho -1\right)}^2}$.
Again, it is not difficult to see that the complex function
$\xi = p\sinh\mu +\imath q \sin\theta$
is a solution of (\ref{b112}) with the
condition $q^2-p^2=1$. Setting $p=\frac{1}{m}\;,\;q=\frac{a}{m}$,
this means  $a^2-m^2=1$.
Equivalently, we can choose
$p=\frac{\sqrt{a^2-m^2}}{m}\;,\;q=\frac{a}{m}$.
From the relation $q^2-p^2=1$ it follows that $p$ 
becomes pure imaginary if $a\rightarrow 0$ and we cannot take the static limit
accordingly with the considerations above.
Concerning the line element
$ds^2=\frac{e^{2\gamma}}{f}\left[d{\mu}^2+d{\theta}^2\right]+
\frac{{\rho}^2}{f}d{\varphi}^2-f{\left(dt-\omega d{\varphi}\right)}^2$,
we get
\begin{eqnarray}
& &f=\frac{p^2{\sinh}^2\mu +q^2{\sin}^2\theta-1}
  {{(p\sinh+1)}^2+q^2{\sin}^2\theta} \; , \; 
\omega=\frac{2q}{p}\frac{(p\sinh\mu+1){\cos}^2\theta}
{[p^2{\sinh}^2\mu-1+q^2{\sin}^2\theta]}\;,\nonumber\\
& &e^{2\gamma}=\frac{p^2{\sinh}^2\mu-1+q^2{\sin}^2\theta}
{p^2}.\label{natale}
\end{eqnarray}  
To cast the metric in the form (\ref{b192bb}) we use the relation
$r=\sqrt{a^2-m^2}\sinh\mu+m$
and  perform the rotation $\theta\rightarrow \theta-\pi/2$. 
We then see that
it is impossible to go from the ``oblate'' solution 
to the ``prolate'' one in a mathematically and 
physically reasonable manner.
\section{Kerr solutions with topological defect}
As a final illustration of the significance of the Papapetrou gauge 
we show how it can be used to study in simple way Kerr solutions in 
the presence of a cosmic string.
All the three metrics we obtained 
(without setting $k^2=p^2$) have, after
rescaling by a factor $p^2/k^2$, a common asymptotic behaviour.
Explicitly (setting G=c=1) 
\begin{equation}
ds^2=d{\rho}^2+dz^2+C^2{\rho}^2d{\varphi}^2-dt^2,
\label{h2}
\end{equation}
where $k^2/p^2=C^2$.
It is a well known fact \cite{wil} that, when $C<1$, 
the space-time (\ref{h2})
is a solution of Einstein's equations with stress-energy tensor
\begin{equation}
T_{\mu\nu}=\mu\delta(x)\delta(y)diag(1,0,0,-1)\;,\;
C=1-4\mu.
\label{h5}
\end{equation}
Expression (\ref{h5}) represents a string
on the $z$ axis with constant mass density $\mu$. 
The parameter $C$ represents a topological defect
with angle deficit $2\pi-2\pi C=8\pi \mu$. 
It is also known \cite{Al} that in the limit 
$\rho\rightarrow 0$ the quantity
\begin{equation}
\Delta\Phi(\rho)=2\pi-
\frac{{\int}^{2\pi}_{0}\sqrt{g_{\varphi\varphi}}d\varphi}
{{\int}^{\rho}_{0}\sqrt{g_{\rho\rho}}d\rho},
\label{top}
\end{equation}
is directly
related to the energy density per unit lengh of the string.
If $\Delta\Phi(0)=0$ the topological defect disappears but
for the three metrics that we are considering  
we have $\Delta\Phi(0)=2\pi(1-C)$. 
Consider first the ``prolate'' case (\ref{b196bb}): 
performing a Taylor expansion in the 
neighbourhood of the $z$ axis ($\rho=0$), we obtain
(outside the horizon)
$\sqrt{g_{\varphi\varphi}}=\frac{\rho\sqrt{p^2 z^2+1+2pz+q^2}}
{p\sqrt{z^2-1}}+o({\rho}^2)$ and
$\sqrt{g_{\rho\rho}}=\frac{\sqrt{p^2 z^2+1+2pz+q^2}}
{Cp\sqrt{z^2-1}}+o(\rho)$. It follows that $\Delta\Phi(0)=2\pi(1-C)=8\pi\mu$.
In a similar way the same result is obtained for the metrics 
(\ref{b516}) and (\ref{natale}).\\
Finally, when the parameter $C$ is used, the metric (\ref{b192bb})
in BL coordinates becomes 
\begin{eqnarray}
& &ds^2 =\frac{\Sigma}{C^2}\left(d{\theta}^2+
\frac{dr^2}{\Delta}\right)+
(r^2+a^2){\sin}^2\theta d{\varphi}^2-dt^2+\nonumber \\
& &+\frac{2mr}{\Sigma}
{\left(dt+a{\sin}^2\theta d\varphi\right)}^2, \nonumber \\
& &\Sigma =r^2+a^2{\cos}^2\theta\;\;\;,\;\;\;\Delta=r^2+a^2-2mr, \label{b616}
\end{eqnarray}
which describes the Kerr solutions with a static string
on the $z$ axis.
\section{Conclusions}
In this paper we have presented a simple novel derivation of equations
appropriate for a stationary axisymmetric space-time using Papapetrou gauge.
The Kerr solutions in this gauge are disconnected: it is 
impossible to go from one to the other by continuously 
changing the parameters. A quantity is introduced which 
represents a topological defect induced
from a static infinitely long cosmic string on the
$z$ axis.
\section*{Acknowledgments}
We would like to thank Roberto Balbinot for his encouragement and
many useful discussions. We would also like to thank Naresh Dadhich
for comments and suggestions.
\section*{APPENDIX}
Using spherical analytic coordinates ($r=e^{\nu}$)
we have the solution
$\xi=p\cos\vartheta+\imath qe^{\nu}$ with $p^2=1$.
For simplicity we take  $p=1$. The metric is
\begin{eqnarray}
& &ds^2=\frac{\left[{\left(1+\cos\vartheta\right)}^2+q^2 r^2\right]}
{k^2 r^2}dr^2+\frac{\left[{\left(1+\cos\vartheta\right)}^2+q^2 r^2\right]}
{k^2}d{\vartheta}^2+ \nonumber \\
& &+r^2{\sin}^2\vartheta\frac{\left[{\left(1+\cos\vartheta\right)}^2+
q^2 r^2\right]}{q^2 r^2-{\sin}^2\vartheta}d{\varphi}^2-\nonumber\\
& &-\frac{q^2 r^2-{\sin}^2\vartheta}{{\left(1+\cos\vartheta\right)}^2+
q^2 r^2}{\left[dt-\frac{2qr^2(1+\cos\vartheta)}
{q^2 r^2-{\sin}^2\vartheta}d\varphi\right]}^2. \label{b711}
\end{eqnarray}
Putting $k=q$, we have in the asymptotic limit
\begin{equation}
ds^2=dr^2+r^2 d{\vartheta}^2+r^2{\sin}^2\vartheta d{\varphi}^2-
{\left[dt-\frac{2}{q}(1+\cos\vartheta)d\varphi\right]}^2.
\label{b713}
\end{equation}
This expression has the same asymptotic behaviour of the N.U.T
solution \cite{NUT}. Recently, the general Kerr-N.U.T. solution
has been obtained (see \cite{DAD}) in a very simple 
form and its uniqueness has
been established for spacetimes admitting
separable equations of motion.

\end{document}